\documentclass[a4paper,fleqn,usenatbib]{mnras}

\usepackage{newtxtext,newtxmath}

\usepackage[T1]{fontenc}
\usepackage{ae,aecompl}

\usepackage{graphicx}	
\usepackage{amsmath}	
\usepackage{nccmath}
\usepackage{amssymb}	
\usepackage{pdflscape}	
\usepackage{multicol}

\title[High-redshift SN~II Hubble diagram]{SN~2016jhj at redshift 0.34: extending the Type II supernova Hubble diagram using the standard candle method}

\author[de Jaeger et al.]
{T. de Jaeger$^{1}$\thanks{E-mail: tdejaeger@berkeley.edu},
L. Galbany$^{2}$, A. V. Filippenko$^{1,3}$, S. Gonz\'{a}lez-Gait\'{a}n$^{4,5}$, N. Yasuda$^{6}$,
\newauthor
K. Maeda$^{6,7}$,
M. Tanaka$^{6,8}$
T. Morokuma$^{6,9}$, 
T. J. Moriya$^{8}$, 
N. Tominaga$^{6,10}$, 
\newauthor
K. Nomoto$^{6}$, 
Y. Komiyama$^{11,12}$,
J. P. Anderson$^{13}$, 
T. G. Brink$^{1}$,
R. G. Carlberg$^{14}$, 
\newauthor
G. Folatelli$^{15,6}$, 
M. Hamuy$^{16,5}$, 
G. Pignata$^{17,5}$,
W. Zheng$^{1}$
\newauthor
\\
\small
$^{1}$Department of Astronomy, University of California, Berkeley, CA 94720-3411, USA.\\
$^{2}$PITT PACC, Department of Physics and Astronomy, University of Pittsburgh, Pittsburgh, PA 15260, USA.\\
$^{3}$Senior Miller Fellow, Miller Institute for Basic Research in Science, University of California, Berkeley, CA 94720, USA.\\
$^{4}$Center for Mathematical Modelling, University of Chile, Beaucheff 851, Santiago, Chile.\\
$^{5}$Millennium Institute of Astrophysics (MAS), Nuncio Monse\~{n}or S\'{o}tero Sanz 100, Providencia, Santiago, Chile.\\
$^{6}$Kavli Institute for the Physics and Mathematics of the Universe (WPI), The University of Tokyo, 5-1-5 Kashiwanoha, Kashiwa, Chiba 277-8583, Japan.\\
$^{7}$Department of Astronomy, Kyoto University, Kitashirakawa-Oiwake-cho, Sakyo-ku, Kyoto, 606-8502, Japan.\\
$^{8}$Division of Theoretical Astronomy, National Astronomical Observatory of Japan, Tokyo 181-8588, Japan.\\
$^{9}$Institute of Astronomy, The University of Tokyo, Mitaka, Tokyo 181-0015, Japan.\\
$^{10}$Department of Physics, Faculty of Science and Engineering, Konan University, 8-9-1 Okamoto, Kobe, Hyogo 658-8501, Japan.\\
$^{11}$National Astronomical Observatory of Japan,2-21-1 Osawa, Mitaka, Tokyo 181-8588, Japan.\\
$^{12}$Graduate University for Advanced Studies (SOKENDAI), 2-21-1 Osawa, Mitaka, Tokyo 181-8588, Japan.\\
$^{13}$European Southern Observatory, Alonso de C\'{o}rdova 3107, Casilla 19, Santiago.\\
$^{14}$Department of Astronomy and Astrophysics, University of Toronto, 50 St. George St., Toronto, ON, M5S 3H4, Canada.\\
$^{15}$Facultad de Ciencias Astron\'{o}micas y Geof\'{i}sicas, UNLP, IALP, CONICET, Paseo del Bosque S/N, B1900FWA La Plata, Argentina.\\ 
$^{16}$Departamento de Astronom\'{i}a -- Universidad de Chile, Camino el Observatorio 1515, Santiago, Chile.\\
$^{17}$Departamento de Ciencias Fisicas, Universidad Andres Bello, Avda. Republica 252, Santiago, Chile.\\
}

\date{Accepted 2017 September 1. Received 2017 September 1; in original form 2017 July 1}

\pubyear{2017}

\begin{document}
\label{firstpage}
\pagerange{\pageref{firstpage}--\pageref{lastpage}}
\maketitle

\begin{abstract}
Although Type Ia supernova cosmology has now reached a mature state, it is important to develop as many independent methods as possible to understand the true nature of dark energy. Recent studies have shown that Type II supernovae (SNe~II) offer such a path and could be used as alternative distance indicators. However, the majority of these studies were unable to extend the Hubble diagram above redshift $z=0.3$ because of observational limitations. Here, we show that we are now ready to move beyond low redshifts and attempt high-redshift ($z \gtrsim 0.3$) SN~II cosmology as a result of new-generation deep surveys such as the Subaru/Hyper Suprime-Cam (HSC) survey. Applying the ``standard candle method'' to SN~2016jhj ($z=0.3398 \pm 0.0002$; discovered by HSC) together with a low-redshift sample, we are able to construct the highest-redshift SN~II Hubble diagram to date with an observed dispersion of 0.27 mag (i.e., 12--13\% in distance). This work demonstrates the bright future of SN~II cosmology in the coming era of large, wide-field surveys like that of the Large Synoptic Survey Telescope.
\end{abstract}

\begin{keywords}
cosmology: distance scale -- galaxies: distances and redshifts -- stars: supernovae: general
\end{keywords}


\section{Introduction}

Type Ia supernovae (SNe~Ia; \citealt{min41,filippenko97}, and references therein) are good standardisable candles (e.g., \citealt{phillips93,hamuy96,riess96,perlmutter97}) that allow measurement of extragalactic distances with an accuracy of $\sim 5\%$ (e.g., \citealt{betoule14,rubin16}). They play a crucial role in the determination of the expansion rate of the Universe \citep[][and references therein]{riess16}, and in 1998 they revealed the surprising accelerated growth rate of the Universe driven by an unknown effect attributed to dark energy \citep{riess98,schmidt98,perlmutter99}. However, even though SNe~Ia remain the most mature and well-exploited method measuring the acceleration, further improvement to constrain the nature of dark energy requires developing as many independent methods as possible.

Among the wide range of independent cosmological probes found in the literature such as the cosmic microwave background radiation (\citealt{fixsen96,jaffe01,spergel07,bennett03,planck13}), baryon acoustic oscillations (\citealt{blake03,seo03}), X-ray clusters (\citealt{white93,schuecker03}), and superluminous SNe (\citealt{inserra14}), one interesting independent method for deriving accurate distances and measuring cosmological parameters is Type II SNe\footnote{Throughout this paper, SNe~II refer collectively to the two historical groups, SNe~IIP and SNe~IIL, since recent studies show that the SN~II family forms a relatively continuous class \citep{anderson14a,sanders15,valenti16,galbany16a}.}. 

Observationally, SNe~II are characterised by the presence of strong hydrogen (H) features in their spectra (see, e.g., \citealt{filippenko00} and \citealt{filippenko97} for overviews), and a plateau of varying steepness in their light curves \citep{barbon79}. Even though detecting SNe~II at high redshift is challenging owing to their relatively low luminosity (1--2 mag fainter than SNe~Ia; \citealt{richardson14}), their use as cosmic distance indicators is motivated by the fact that they are more abundant than SNe~Ia \citep{li2011}; additionally, their rate is expected to peak at higher redshifts than that of SNe~Ia \citep{taylor14,cappellaro15}. Also, thanks to direct progenitor detections and hydrodynamical models \citep{vandyk03,smartt09a,grassberg71,falk77,chevalier76}, their progenitors and environments (only late-type galaxies) are better understood than those of SNe~Ia (no direct progenitor detection and found also in elliptical galaxies, not just late-type galaxies). It is now accepted that their progenitors are red supergiants that have retained a significant fraction of their H envelopes. Unlike SNe~Ia, for which possible redshift evolution is debated \citep{kessler09,guy10,betoule14}, SN~II~progenitors have been constrained, and the explosion mechanism is better understood \citep{woosley95,janka07}.

At first sight, the SN~II family displays a large range of peak luminosities; however, as for SNe~Ia, their extrinsic differences such as dust extinction can be calibrated. To date, several methods have been developed to standardise SNe~II, such as the expanding photosphere method (EPM; \citealt{kirshner74}), the standard candle method (SCM; \citealt{hamuy02}), the photospheric magnitude method (PMM; \citealt{rodriguez14}), and the most recent technique based solely on photometric inputs called the photometric colour method (PCM; \citealt{dejaeger15b,dejaeger17a}). In this paper, we focus our effort on the SCM, which is the most common method used to derive SN~II distances, and currently the most accurate. The SCM is a powerful method based on both photometric and spectroscopic input parameters which enables a decrease of the scatter in the Hubble diagram from $\sim$ 1 mag (with calibration) to levels of $\sim$ 0.3 mag \citep{hamuy02}, equivalent to a precision of $\sim $14\% in distances. This method is mainly built on the observed correlation between SN~II luminosity and photospheric expansion velocity $\sim$ 50 days post-explosion: more luminous SNe~II have higher velocities \citep{hamuy02}. The underlying physical cause of this empirical relation \citep{kasen09} is that in more-luminous SNe~II the H recombination front defining the photosphere is farther out in radius, and therefore maintained at higher velocities because of homologous expansion. 

Currently, refined versions of the SCM combine the velocity ejecta correction measured through the absorption minimum of P-Cygni features (\ion{Fe}{II} $\lambda 5169$ or H$\beta$ $\lambda$4861 lines) with a dust correction based on the SN~II colour. Using this fine-tuned method, recent works succeeded in constructing a Hubble diagram with a dispersion of $\sim10$--12\% in distance \citep{nugent06,poznanski09,olivares10,andrea10,dejaeger15b,dejaeger17a}, and provide new independent evidence for dark energy at the level of 2$\sigma$ \citep{dejaeger17a}. Even if SNe~II are currently not competitive with SNe~Ia in terms of achieved dispersion ($\sim$ 0.25 mag versus $\sim$ 0.10 mag) or sample size ($\sim$ 60 SNe~II versus $\sim$ 740 SNe~Ia), the latest SN~II studies are comparable to the early SN~Ia results, showing that SNe~II are a useful complementary and independent method to constrain the nature of dark energy.

Nevertheless, all SN~II Hubble diagrams based on the SCM found in the literature used relatively low-redshift samples ($z \lesssim 0.2$), except the recent work done by \citet{gall17} where the authors used one SN~II at a redshift of 0.335. At low redshift, the differences between the expansion histories are extremely small, and distinguishing among the different cosmological models requires measurements extending far back in time: SNe~II at higher redshift (at least $z\approx 0.3$--0.5). Here, as proof of concept, we show our ability to extend the current SN~II Hubble diagram beyond $z=0.3$, taking advantage of the Subaru/Hyper Suprime-Cam survey (HSC; \citealt{miyazaki12,aihara17a}) and using different methodologies than those used by \citet{gall17}. 

This work also addresses the critical issue of the necessity for the community to dedicate more observations of high-redshift SNe~II in order to directly compare with SN~Ia results. This could be achieved with new, deep surveys (e.g., the Large Synoptic Survey Telescope, LSST: \citealt{ivezic09}; HSC: \citealt{miyazaki12,aihara17a}) and ground-base telescopes for spectroscopy such as the Keck telescopes or the next generation of 25--39~m telescopes (European Extremely Large Telescope, E-ELT: \citealt{gilmozzi07}; Giant Magellan Telescope, GMT: \citealt{johns12}; Thirty Meter Telescope, TMT: \citealt{sanders13}).

This paper is organised as follows. Section 2 contains a description of the data sample, and in Section 3 we discuss the method used to derive the Hubble diagram, which differs from that of \citet{dejaeger17a}. In Section 4 we discuss our results, and Section 5 summarised our conclusions.

\section{Data Sample}

Here we use the sample from \citet{dejaeger17a}, which consist of SNe~II from three different projects: the Carnegie Supernova Project\footnote{\url{http://csp.obs.carnegiescience.edu/}} (CSP-I; \citealt{ham06}), the SDSS-II SN Survey\footnote{\url{http://classic.sdss.org/supernova/aboutsupernova.html}} \citep{frieman08}, and the Supernova Legacy Survey\footnote{\url{http://cfht.hawaii.edu/SNLS/}} \citep{astier06,perrett10}. We complete the sample with SN 2016jhj, a recently discovered high-redshift SN~II from the HSC (\citealt{aihara17a}).

\subsection{CSP-I+SDSS+SNLS}

A list of 82 SNe~II available for the SCM was compiled by \citet{dejaeger17a}. Among this sample, 61 low-redshift SNe~II are from the CSP-I (Anderson et al., in prep.), 16 from the SDSS \citep{andrea10}, and 5 from the SNLS (unpublished data). For more information about the different surveys and data-reduction procedures, the reader is referred to \citet{andrea10}, \citet{dejaeger17a}, and references therein.

All of the magnitudes were simultaneously corrected for Milky Way extinction ($A_V$G; \citealt{schlafly11}), the expansion of the Universe (K-correction; \citealt{oke68,hamuy93,kim96,nugent02}), and differences between the photometric systems (S-correction; \citealt{stritzinger02}) using the cross-filter K-corrections defined by \citet{kim96}. The procedures are described in detail by \citet{nugent02}, \citet{hsiao07}, and \citet{dejaeger17a}. Note that this correction will be hereafter referred to as the AKS correction.

\subsection{Subaru/HSC}

To extend the Hubble diagram beyond $z=0.3$, we use data from the HSC
installed at the prime focus of the 8.2~m Subaru telescope, and the Low Resolution Imaging Spectrometer (LRIS; \citealt{oke95}, \citealt{rockosi10}) on the Keck-I 10~m telescope, both located at the Mauna Kea Observatory.
HSC is a new instrument with the currently highest etendue, a combination of field of view (1.5$^{\circ}$ in diameter) and telescope size (8.2~m). The HSC transient survey started in November 2016 and had guaranteed
access to 5--7 nights per month (dark phase) during six months. For each run, the COSMOS field \citep{capak07} is observed about twice in each filter ($grizy$). The data are reduced with \textsc{hscPipe} \citep{bosch17}, a version of the LSST stack \citep{ivezic09,alexrod10,juric15}. HSC astrometry and photometry are calibrated relative to the Pan-STARRS~1 (PS1) 3$\pi$ catalog \citep{schlafly12,tonry12,magnier13}. Final photometric points are obtained via point-spread-function photometry in the template-substracted images. We refer the reader to \citet{aihara17a} for more detailed information regarding photometric reduction and Yasuda et al. (in prep.) for details of the COSMOS transient survey.

\begin{figure}
\includegraphics[width=1.0\columnwidth]{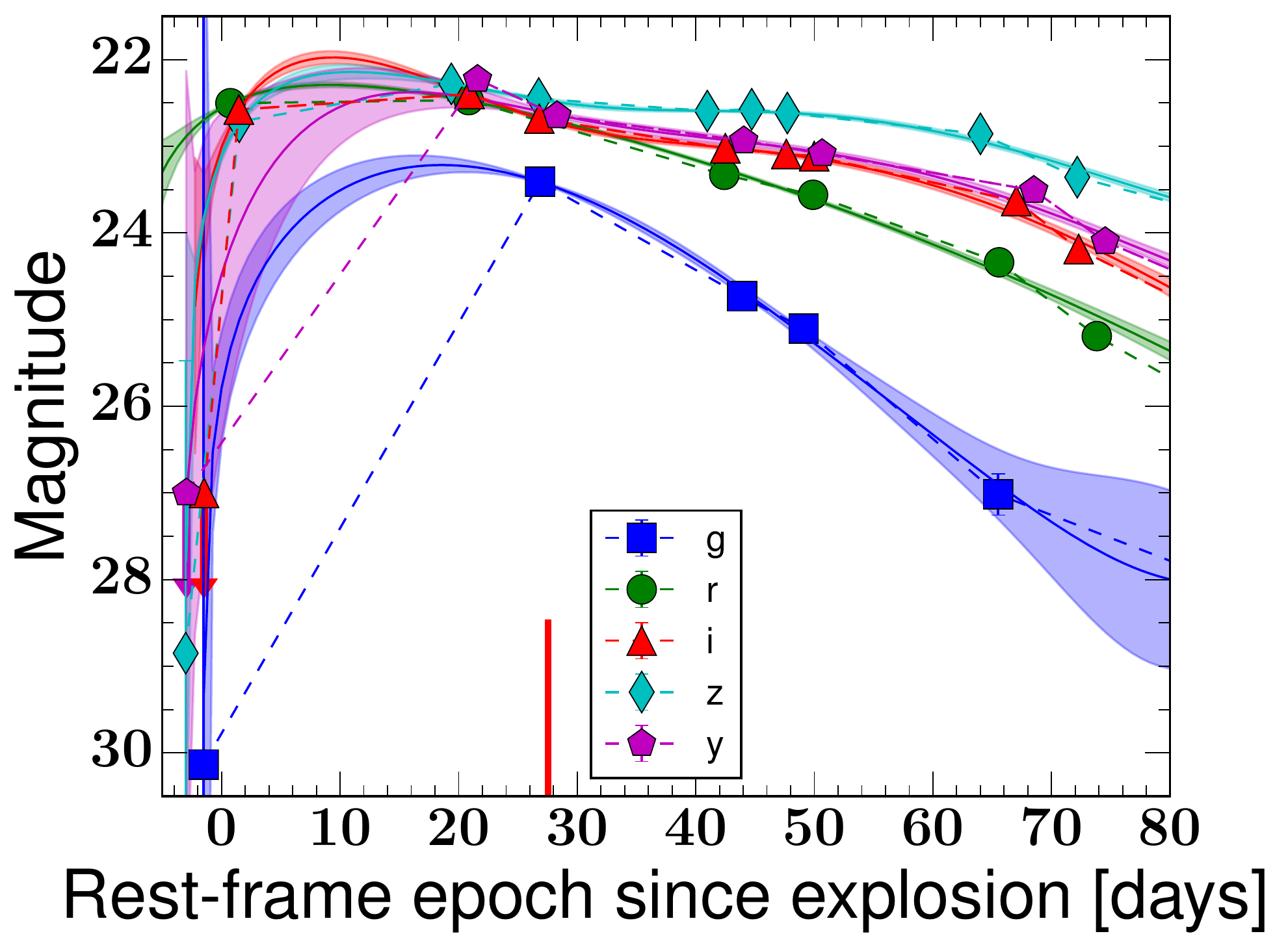}
\caption{SN~2016jhj $grizy$ light curves from the Hyper Suprime-Cam on the 8.2~m Subaru telescope. The blue squares, green circles, red triangles, cyan diamonds, and magenta pentagons are (respectively) the $g$, $r$, $i$, $z$, and $y$ light curves. The vertical red line represents the epoch of the visual-wavelength spectrum since the explosion. For each filter, the solid curve is the best fit from the Gaussian-process interpolation described in Section \ref{section_GP}. The filled region represents the 1$\sigma$ uncertainty of the regression curve. The dashed lines are shown only to guide the eye. The assumed explosion date is JD 2,457,716.6 $\pm$ 1.0.}
\label{fig:sn2016jhj_photo}
\end{figure}

When a SN candidate is detected, a photometric selection is performed based on two criteria specific to SNe~II: a rapid light-curve rise \citep{gonzalezgaitan14} and a plateau resulting from the balance between cooling and recombination in the ejecta \citep{grassberg71}. In Figure \ref{fig:sn2016jhj_photo}, the $grizy$ light curves of SN~2016jhj are displayed and the photometry is given in Table \ref{tab:sn2016jhj_photo}. As Figure \ref{fig:sn2016jhj_photo} shows, the first SN~II criterion is satisfied with a fast light-curve rise time in the $z$ band of $\sim$ 5 days. The second SN~II criterion is also fulfilled with the presence of a clear plateau in the $rizy$ light curves at an apparent magnitude of $\sim$ 23 mag for more than 50 days. To estimate the explosion date we use the $i$ band, for which the first photometric point (JD 2,457,721.55; Nov. 29, 2016; UT dates are used throughout this paper) was obtained four days after nondetection (JD 2,457,717.62; Nov. 25, 2016). In this paper, the explosion date is chosen as the median value between these two epochs and thus is estimated to be JD 2,457,719.56 $\pm$ 2. It is worth noting the quality of the light curve obtained using the HSC relative to one recent high-redshift SN~II (PS1-13bni; \citealt{gall17}) observed by the Panoramic Survey Telescope \& Rapid Response System 1 (Pan-STARRS1; \citealt{kaiser10}).
 
\begin{table}\small
\caption{Host-galaxy and SN parameters.}
\centering
\begin{tabular}{llc}
\hline
\hline
Parameters & Values & Refs \\
\hline
\textbf{Host galaxy:} & & This work\\
Name & Anonymous &--\\
Type & Unknown &--\\
RA (J2000) &$\alpha = 09\fh 58\fm 53\fs485$ &--\\
Dec (J2000) & $\delta= +02^{\circ}01{\farcm}28{\farcs}62$ &--\\
Heliocentric redshift &$z_\mathrm{hel}=0.3398 \pm 0.0002$ &--\\
\textbf{Supernova:} & & This work\\
Name & SN~2016jhj & --\\
Type & SN~II & --\\
RA (J2000) &$\alpha = 09\fh 58\fm 53\fs505$ &--\\
Dec (J2000) & $\delta= +02^{\circ}01{\farcm}28{\farcs}45$ &--\\
Explosion date (JD) & 2,457,719.6 $\pm 2$ &--\\
Milky Way extinction (mag) & $E(B-V)=0.0515$ & (1)\\
Distance modulus (mag) & $\mu$= 41.07 $\pm$ 0.12$^{2}$ $\pm$ 0.27$^{3}$ &--\\
\hline
\end{tabular}
Note: (1) \citet{schlafly11}; (2) uncertainty accounted for measurement errors of SN~2016jhj (velocity, magnitude, colour, redshift) of SN~2016jhj; (3) observed dispersion from the SN~II Hubble diagram ($\sigma_\mathrm{obs}$).
\label{table:sn_parameters}
\end{table}

\begin{table}\small
\caption{SN~2016jhj $grizy$ fluxes and magnitudes, and $1\sigma$ uncertainties.}
\centering
\begin{tabular}{lcccc}
\hline
\hline
JD $-$ & Filters & flux (err) & mag (err) \\
2,400,000 &  &  & \\
\hline
57717.57  &HSC-$g$    &0.056    (0.560)&  30.134  (10.908)\\
57755.61  &HSC-$g$   &27.298    (0.208)&  23.410   (0.008)\\
57778.45  &HSC-$g$    &8.077    (0.241)&  24.732   (0.032)\\
57785.39  &HSC-$g$    &5.738    (0.211)&  25.103   (0.040)\\
57807.37  &HSC-$g$    &0.983    (0.215)&  27.018   (0.238)\\
57834.32  &HSC-$g$    &0.370    (0.193)&  28.080   (0.568)\\
57841.29  &HSC-$g$    &0.428    (0.202)&  27.921   (0.513)\\
57869.33  &HSC-$g$    &0.269    (0.265)&  28.425   (1.068)\\
57720.60 &HSC-$r2$   &62.929    (0.251)&  22.503   (0.004)\\
57747.53 &HSC-$r2$   &64.971    (0.353)&  22.468   (0.006)\\
57776.42 &HSC-$r2$   &29.365    (0.280)&  23.330   (0.010)\\
57786.45 &HSC-$r2$   &23.778    (0.218)&  23.560   (0.010)\\
57807.48 &HSC-$r2$   &11.595    (0.282)&  24.339   (0.026)\\
57818.52 &HSC-$r2$    &5.289    (0.274)&  25.191   (0.056)\\
57837.26 &HSC-$r2$    &1.938    (0.292)&  26.282   (0.164)\\
57844.33 &HSC-$r2$    &2.441    (0.245)&  26.031   (0.109)\\
57866.25 &HSC-$r2$    &1.767    (0.327)&  26.382   (0.201)\\
57717.62 &HSC-$i2$   &-0.577    (0.352)&   0.000   (0.000)\\
57721.55 &HSC-$i2$   &58.780    (0.432)&  22.577   (0.008)\\
57747.62 &HSC-$i2$   &69.314    (0.404)&  22.398   (0.006)\\
57755.51 &HSC-$i2$   &53.383    (0.290)&  22.681   (0.006)\\
57776.54 &HSC-$i2$   &38.952    (0.279)&  23.024   (0.008)\\
57783.43 &HSC-$i2$   &36.641    (0.361)&  23.090   (0.011)\\
57786.58 &HSC-$i2$   &35.821    (0.521)&  23.115   (0.016)\\
57809.42 &HSC-$i2$   &22.269    (0.472)&  23.631   (0.023)\\
57816.47 &HSC-$i2$   &13.340    (0.262)&  24.187   (0.021)\\
57835.26 &HSC-$i2$   & 5.563    (0.427)&  25.137   (0.083)\\
57842.27 &HSC-$i2$   & 5.437    (0.426)&  25.162   (0.085)\\
57869.27 &HSC-$i2$   & 3.707    (0.543)&  25.578   (0.159)\\
57870.35 &HSC-$i2$   & 5.769    (0.628)&  25.097   (0.118)\\
57715.55  &HSC-$z$   & 0.182    (0.566)&  28.849   (3.372)\\
57721.60  &HSC-$z$   &51.651    (0.621)&  22.717   (0.013)\\
57745.57  &HSC-$z$   &77.277    (0.637)&  22.280   (0.009)\\
57755.45  &HSC-$z$   &66.068    (0.610)&  22.450   (0.010)\\
57774.50  &HSC-$z$   &57.847    (0.255)&  22.594   (0.005)\\
57779.52  &HSC-$z$   &59.008    (2.391)&  22.573   (0.044)\\
57783.55  &HSC-$z$   &56.668    (0.449)&  22.617   (0.009)\\
57805.37  &HSC-$z$   &45.425    (0.571)&  22.857   (0.014)\\
57816.31  &HSC-$z$   &28.646    (0.491)&  23.357   (0.019)\\
57834.44  &HSC-$z$   &18.206    (0.457)&  23.849   (0.027)\\
57841.41  &HSC-$z$   &18.320    (0.582)&  23.843   (0.034)\\
57866.36  &HSC-$z$   &13.589    (0.887)&  24.167   (0.071)\\
57872.26  &HSC-$z$   &13.230    (0.740)&  24.196   (0.061)\\
57924.28  &HSC-$z$   & 5.033    (2.587)&  25.246   (0.558)\\
57715.62  &HSC-$y$   &-0.732    (0.771)&   0.000   (0.000)\\
57748.53  &HSC-$y$   &81.048    (7.328)&  22.228   (0.098)\\
57757.52  &HSC-$y$   &55.130    (1.238)&  22.647   (0.024)\\
57778.62  &HSC-$y$   &42.227    (2.034)&  22.936   (0.052)\\
57787.47  &HSC-$y$   &36.858    (0.880)&  23.084   (0.026)\\
57811.40  &HSC-$y$   &24.922    (1.364)&  23.509   (0.059)\\
57819.47  &HSC-$y$   &14.463    (0.751)&  24.099   (0.056)\\
57833.38  &HSC-$y$   & 8.264    (1.115)&  24.707   (0.147)\\
57837.49  &HSC-$y$   &16.558   (10.402)&  23.952   (0.682)\\
57842.34  &HSC-$y$   & 8.474    (1.266)&  24.680   (0.162)\\
57863.28  &HSC-$y$   & 8.128    (1.571)&  24.725   (0.210)\\
\hline
\end{tabular}
\label{tab:sn2016jhj_photo}
\end{table}

\begin{figure*}
\includegraphics[width=2.0\columnwidth]{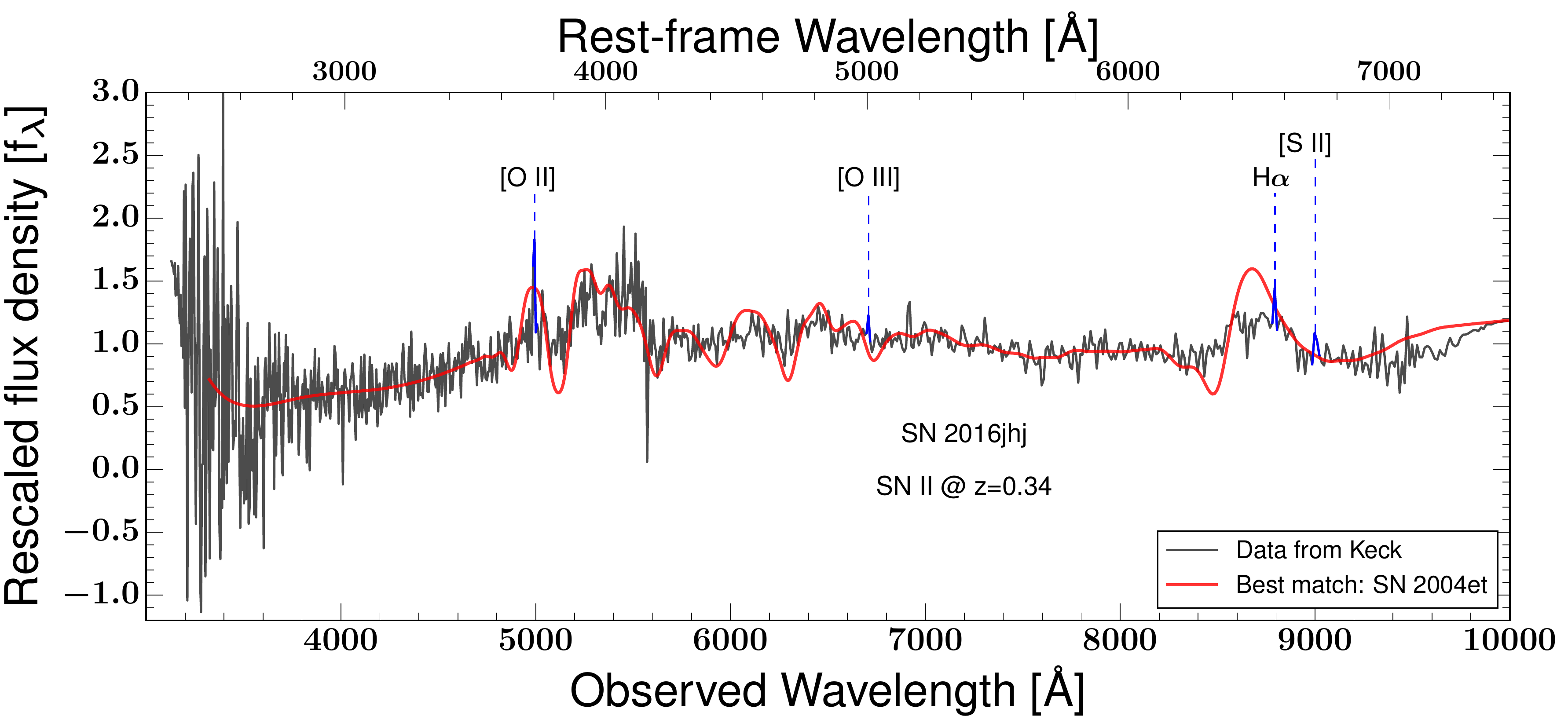}
\caption{SN~2016jhj visual-wavelength spectrum obtained with LRIS on the Keck-I telescope. The SN~2016jhj spectrum (in black) is compared to a SN spectral library (in red) using SNID. The best fit is SN~2004et, a SN~II. Note that the spectrum is binned on a logarithmic wavelength axis \citep{blondin07} and the redshift written was derived from the narrow host-galaxy emission lines. Four emission lines from the host galaxy are also highlighted in blue: [O~II] $\lambda$3727.3, [O~III] $\lambda$5006.8, H$\alpha$ $\lambda$6562.81, and [S~II] $\lambda$6717.67.}
\label{fig:sn2016jhj_spec}
\end{figure*}

From this first HSC run, SN~2016jhj was our best high-redshift SN~II candidate, with an estimated host-galaxy photometric redshift of $\sim 0.33$. Details of the host-galaxy and SN~2016jhj are listed in Table \ref{table:sn_parameters}. As a proof of concept to demonstrate that we are now able to attempt high-redshift SN~II cosmology ($z \gtrsim 0.3$), we obtained an optical spectrum using LRIS on the Keck-I telescope on 2017 January 3 (JD 2,457,756; 29.40 days after the explosion in the rest frame), with a total exposure time of one hour. LRIS is equipped with an atmospheric dispersion corrector, so the correct continuum shape is obtained. The reduction and the calibration were performed following standard procedures (bias subtraction, flat-field correction, one-dimensional extraction, wavelength and flux calibration) described by \citet{silverman12}. In Figure \ref{fig:sn2016jhj_spec}, the observed spectrum (in black) together with the best SN spectral library match found (in red) by the Supernova Identification code (SNID; \citealt{blondin07}) are shown. The best match corresponds to the Type II SN~2004et \citep{sahu06}, which confirms the nature of our transient. Note that among all matches found by SNID (70), more than 85\% were SNe~II and around 15\% SNe~IIb (e.g., SN~1993J; \citealt{filippenko93}). However, it is worth noting that SNe~IIb can be clearly excluded based on the observed light curve.

The heliocentric redshift used in this work was derived directly from the SN spectrum using the superimposed host-galaxy emission lines. In the spectrum, we are able to identify three features: [O~II] $\lambda$3727.3, [O~III] $\lambda$5006.8, and H$\alpha$ $\lambda$6562.81, for which we respectively obtain very consistent redshifts of $0.3397 \pm 0.0001$, $0.3399 \pm 0.0002$, and $0.3400 \pm 0.0001$ (note also the possible presence of the [S~II] $\lambda$6717 emission line at the same redshift).In Figure \ref{fig:sn2016jhj_spec}, the emission lines used to derive the redshift are highlighted in blue. The final heliocentric redshift value is taken as $0.3398 \pm 0.0002$ (weighted average). Finally, owing to a lack of telescope time, we were able to observe only one candidate, even though the HSC survey provided us with additional SN~II candidates.

\section{Methodology}

We now describe how we derive the Hubble diagram, and in particular how the expansion velocities, magnitudes, and colours are estimated. In this paper, the methodology and SN~2016jhj are the main differences with \citet{dejaeger17a}.

\subsection{Expansion velocities}\label{txt:cc_method}

To apply the SCM, it is fundamental to measure the expansion velocity of the ejecta. Generally, the velocity is estimated from the minimum flux of the absorption component of the P-Cygni profile of an optically thin line formed by pure scattering, such as \ion{Fe}{ii} $\lambda$5018. However, for noisy spectra and at early epochs, the \ion{Fe}{ii} features are often buried in the noise, making them unsuitable for velocity measurement. Even though it is less associated with the photospheric velocity, more recently the H$\beta$ $\lambda$4861 absorption line has been proposed \citep{nugent06,poznanski10,takats12,dejaeger17a} as a velocity proxy. This line has the advantage of being stronger than \ion{Fe}{II} $\lambda$5018, and thus easier to measure. Note that a linear relation was derived between the H$\beta$ and \ion{Fe}{II} velocities \citep{poznanski10,takats12,dejaeger17a}, with a dispersion equivalent to an additional uncertainty of about 200--400 km s$^{-1}$.

However, at high redshift, spectra are so noisy that even the H$\beta$ $\lambda$4861 absorption is not clearly visible; hence, some authors attempt to use other methods. In this section, we describe the cross-correlation technique developed by \citet{poznanski09}.

\begin{figure}
\includegraphics[width=1.0\columnwidth]{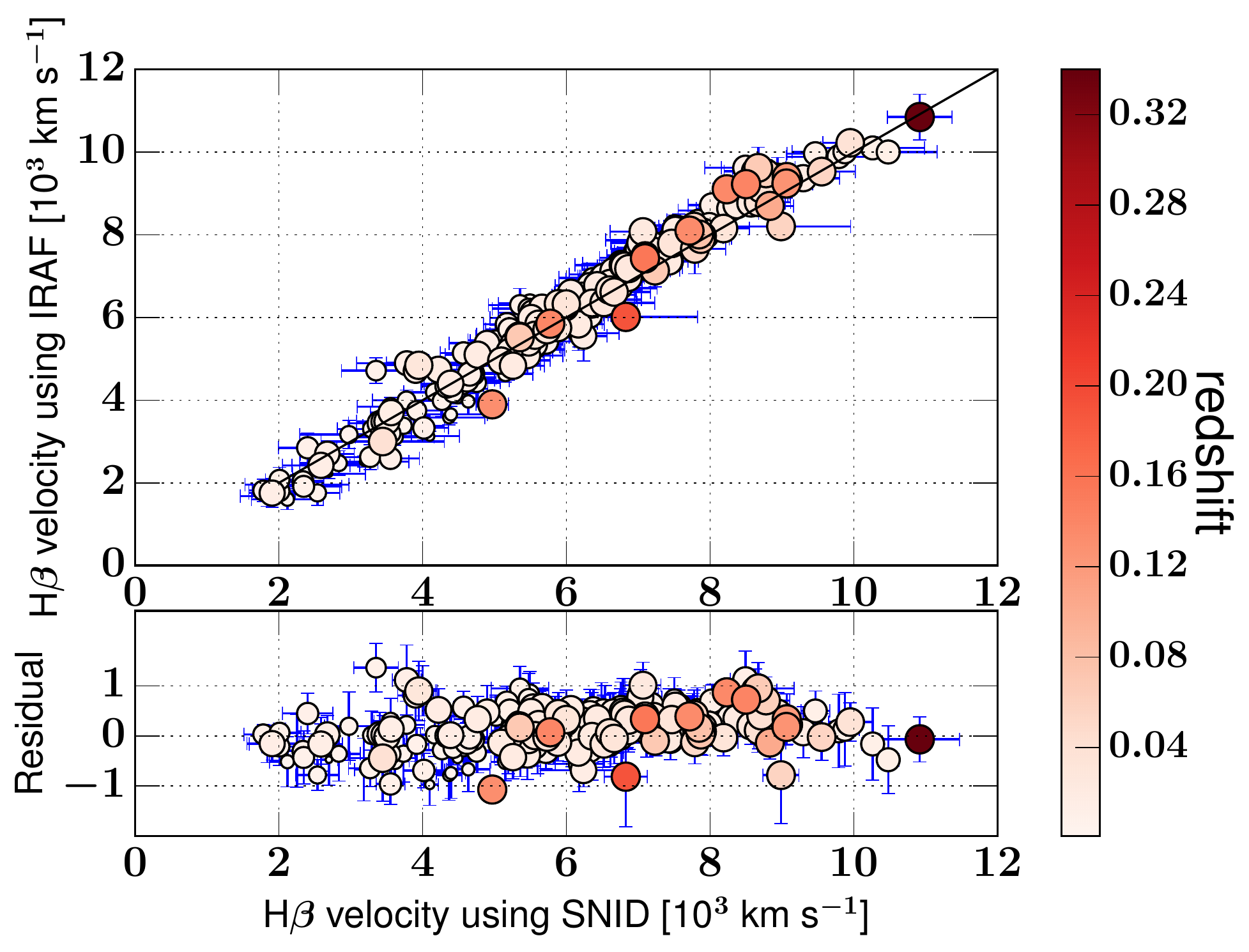}
\caption{Comparison between the H$\beta$ $\lambda$4861 velocities measured from the cross-correlation method with SNID and those determined from the absorption minima of H$\beta$ $\lambda$4861 using IRAF. The residuals are plotted in the bottom panel. The black line represents a slope of unity. The colour bar on the right side represents the different redshifts. The marker size increases with redshift in order to highlight the high-redshift spectra --- those typically with more noise. Note that the brown point at top right corresponds to SN 2016jhj, our distant SN~II.}
\label{fig:SNID_IRAF}
\end{figure}

In this method, the H$\beta$ $\lambda$4861 velocity is determined by computing the cross-correlation (using the \citealt{tonry79} algorithm) between the observed spectra and a library of SNe~II using SNID for which the H$\beta$ $\lambda$4861 velocities can be measured precisely. This method has several advantages such as its simplicity, its robustness, and its utility for broad and noisy lines.

We used templates of six SNe~II from the original SNID library together with a new set of spectral templates (Guti\'errez et al., in prep.). This new library consists of SNe~II with well-constrained explosion epochs from the CSP-I and the Carnegie Type II Supernova Survey \citep{galbany16a}. Briefly, we cross-correlated each observed spectrum only to the SN~II library, constraining the wavelength range to 4400--6000\AA\ (rest frame) in order to avoid the H$\alpha$ $\lambda$6563 line. As the redshift is known, we use the ``forcez'' option available in SNID to directly run a set of correlations, with the input and template spectra trimmed to match at this redshift. For each spectrum, the resulting velocities are the sum of the template velocities and the relative Doppler shift between the observed spectrum and the template. Finally, we select only the velocities of the best-fitting templates (top 10\% of $rlap$\footnote{The $rlap$ parameter is akin to a quality parameter: the higher the $rlap$, the better the correlation \citep{blondin07}.} values) and calculate a weighted velocity mean. Note also that the spectrum epoch is not used as an input parameter because SNe~II evolve at different speeds; we employ only the observed spectra with epochs between 15 and 90 days post-explosion, during the photospheric phase. In order to test this method, for each spectrum the H$\beta$ $\lambda$4861 velocity is also estimated through the minimum flux of the absorption component of the P-Cygni profile using IRAF\footnote{IRAF is distributed by the National Optical Astronomy Observatory, which is operated by the Association of Universities for Research in Astronomy (AURA) under cooperative agreement with the US National Science Foundation (NSF).}. In Figure \ref{fig:SNID_IRAF}, the H$\beta$ $\lambda$4861 velocities measured from both methods are compared. As we can see, the velocities derived from the minimum flux of the H$\beta$ $\lambda$4861 absorption line and those estimated using SNID are correlated with a dispersion of $\sim 400$ km s$^{-1}$. The correlation is also good for the high-$z$ SN~II spectra with a lower signal-to-noise ratio (S/N). This exercise validates the use of the cross-correlation method presented by \citet{poznanski09} and shows its importance for the high-$z$ spectrum velocity measurement. However, it is worth remarking that the cross-correlation method introduces a bias, as we clearly see a trend in the residuals (Pearson factor of $\sim0.35$). As already mentioned by \citet{andrea10} and \citet{takats12}, this bias is caused by the template selection, and it is more important at early and late epochs. Note also that SN~2016jhj has the highest velocity among the the sample, merely due to a Malmquist bias: as the HSC survey searches for the highest-redshift SNe, they then to find only very luminous objects.\\

\indent After computing the H$\beta$ $\lambda$4861 velocity for each SN at all available epochs, we do an interpolation/extrapolation using a power law of the form \citep{hamuyphd}
\begin{ceqn}
\begin{align}
v(t) = A t^{\gamma},
\label{velocity}
\end{align}
\end{ceqn}
where $A$ and $\gamma$ are two free parameters obtained by least-squares minimisation for each individual SN and $t$ is the rest-frame epoch from the explosion. Following the work done by \citet{dejaeger17a}, a Monte Carlo simulation is performed to obtain the velocity uncertainty. We also add to this error a value of 150 km s$^{-1}$, in quadrature, to account for unknown host-galaxy peculiar velocities. When a SN has only one spectrum available, the free parameter $\gamma$ is fixed to a value of $-0.407 \pm 0.173$, which corresponds to the median value derived from the CSP sample \citep{dejaeger17a}. Note the difference between the methodology used by \citet{gall17} and in this work: \citet{gall17} measured the H$\beta$ $\lambda$4861 velocity using IRAF, which could be very difficult for very high-redshift SNe~II and their noisy spectra. Additionally, to derive the velocity at 50 days post-explosion, they assumed the same power law for each SN, while in our work, we derive a specific power law for each SN having at least two spectra.

\begin{figure}
\includegraphics[width=1.0\columnwidth]{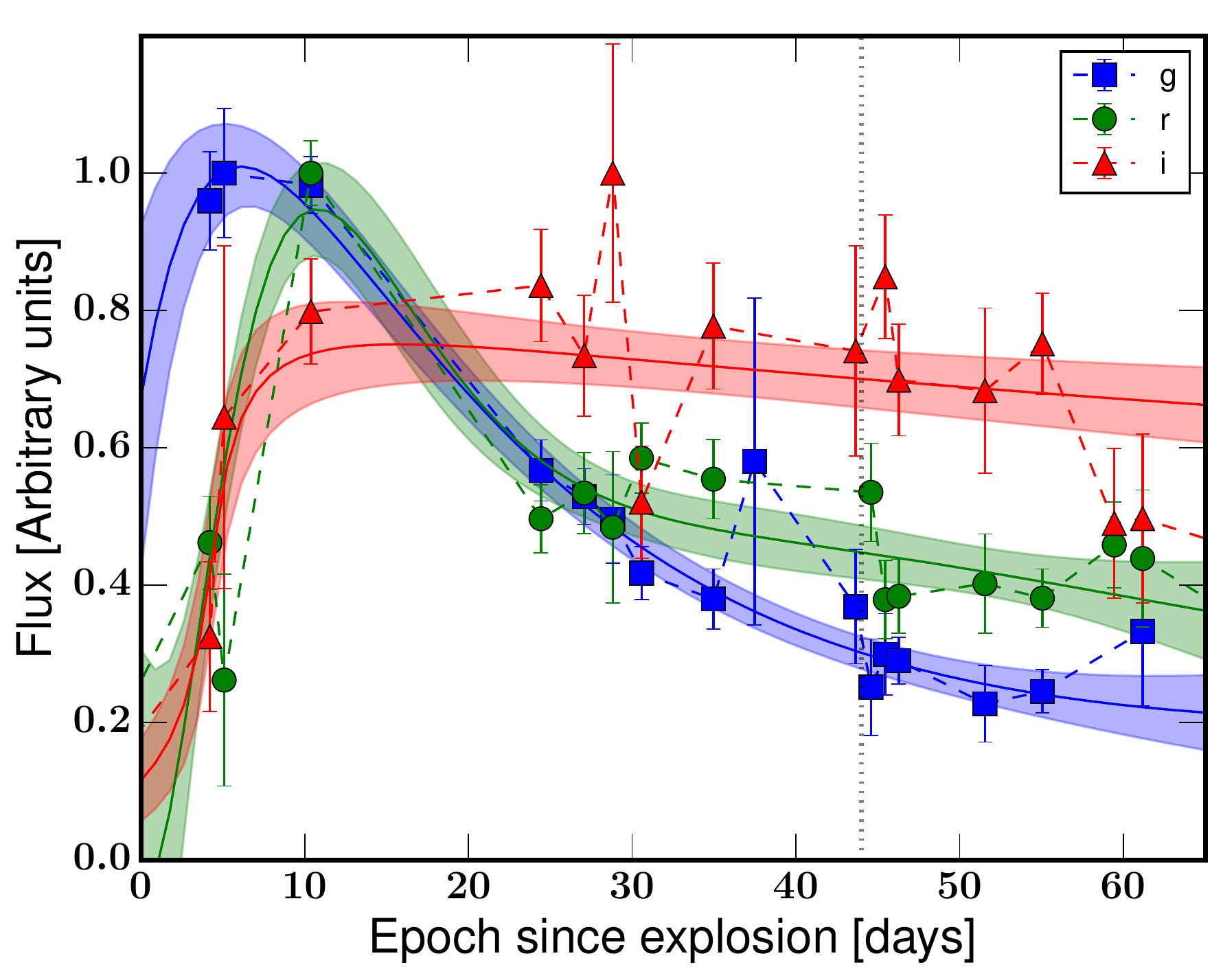}\\
\includegraphics[width=1.03\columnwidth]{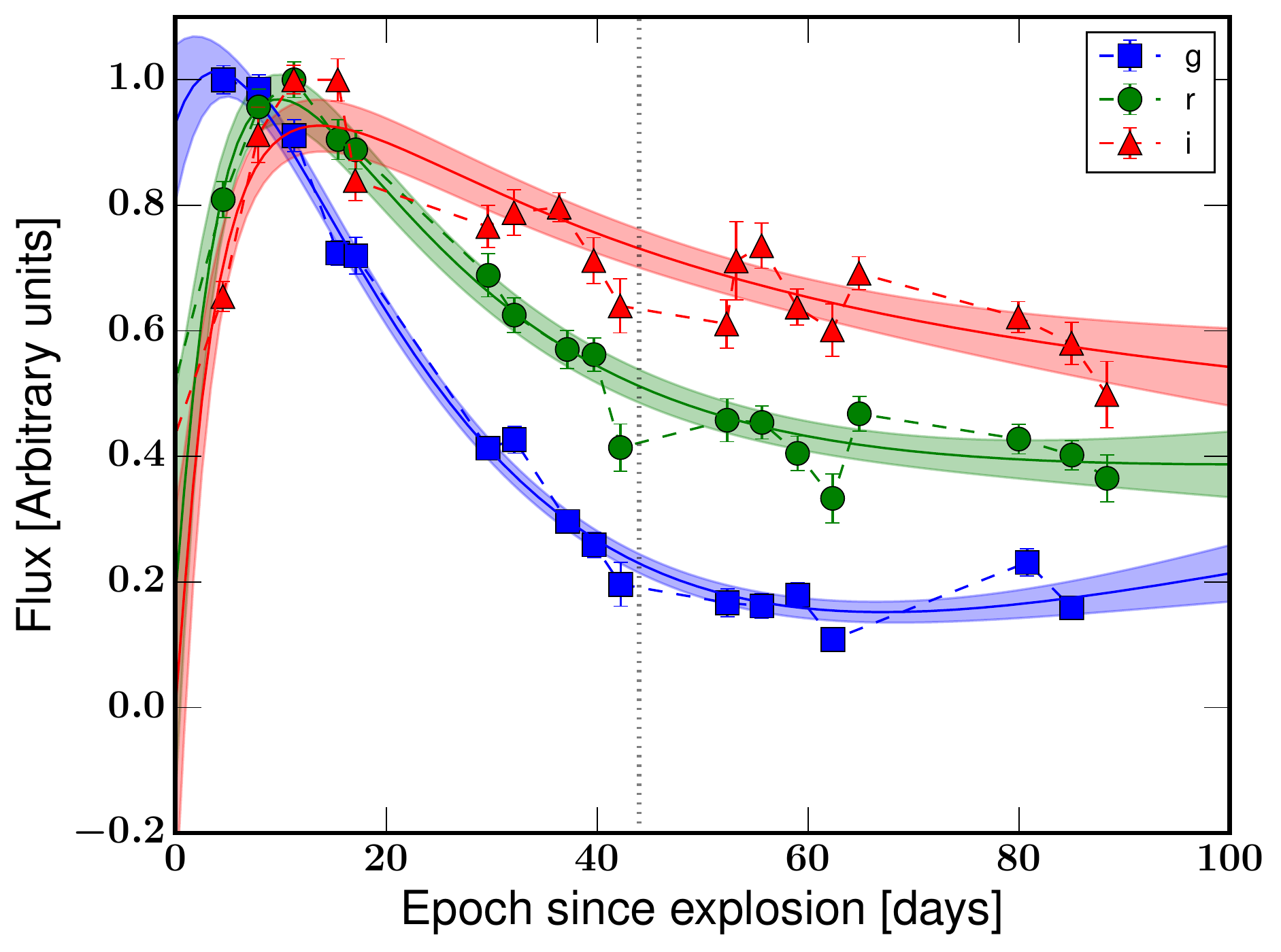}
\caption{\textit{Top:} SN~2007nr AKS-corrected light curves in flux. The explosion date and heliocentric redshift are (respectively) JD 2,454,542.9 $\pm$ 6.0 and $z=0.14$. \textit{Bottom:} SN~05D4dn AKS-corrected light curves in flux. The explosion date and heliocentric redshift are JD 2,454,742.7 $\pm$ 9.0 and $z=0.19$, respectively. For both SNe, the $gri$ bands are shown in blue, red, and green (respectively). For each filter, the solid curve is the best fit from the Gaussian-process regression. The filled region represents the 1$\sigma$ uncertainty of the regression curve. The dotted black line represents the epoch 44 days after the explosion in order to compare between the linear and the Gaussian-process interpolation.}
\label{fig:GP}
\end{figure}

\begin{figure*}
\centering
\includegraphics[width=0.8\textwidth]{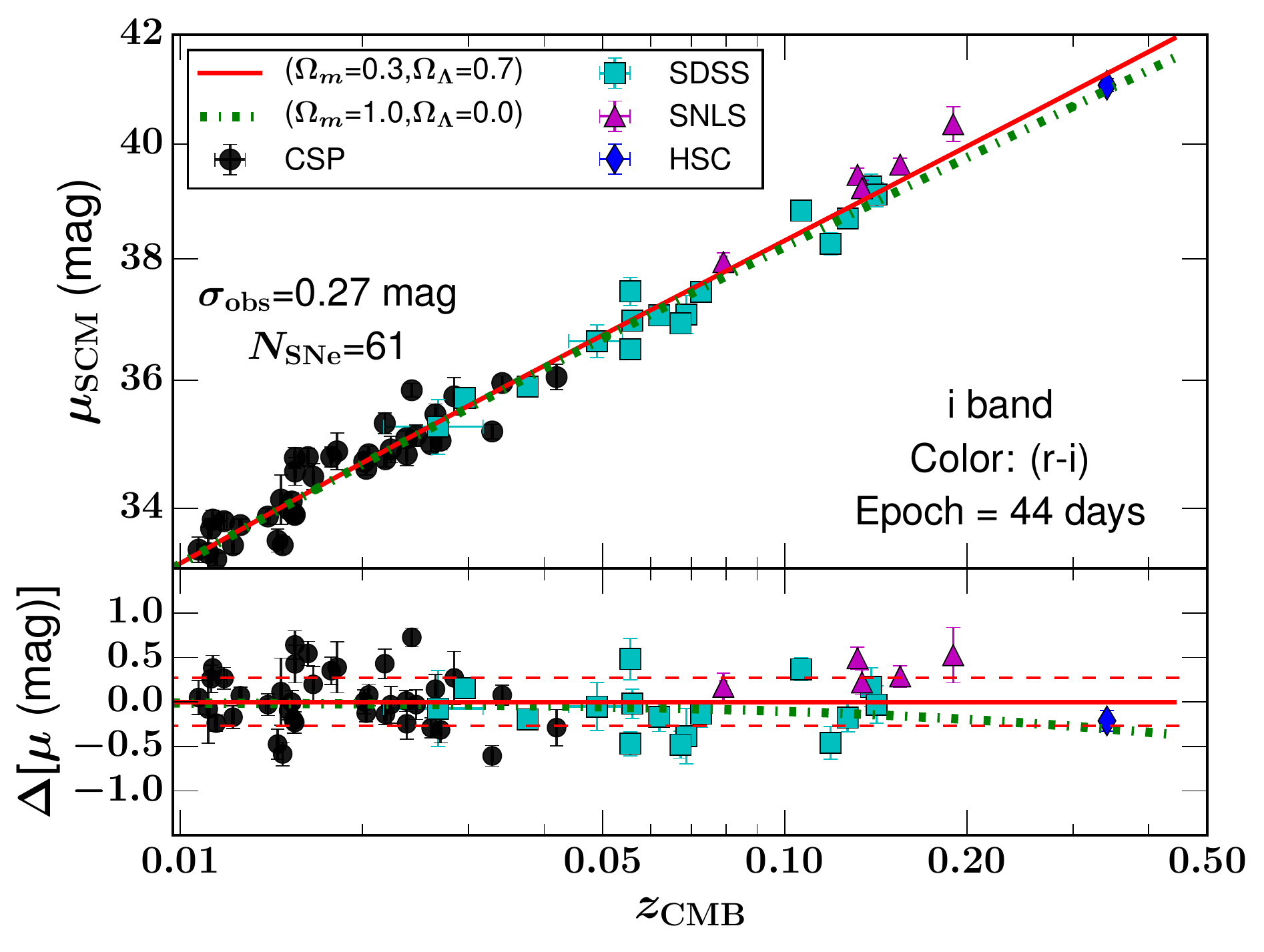}
\caption{Hubble diagram (top) and residuals from the $\Lambda$CDM model (bottom) using the SCM as applied to the data taken from CSP-I (black circles; de Jaeger et al. 2017), SDSS (cyan squares; D'Andrea et al. 2010), SNLS (magenta triangles; de Jaeger et al. 2017), and HSC (blue diamond; this work). The red solid line is the Hubble diagram for the $\Lambda$CDM model ($\Omega_{m}=0.3$, $\Omega_{\Lambda} = 0.7$), while the green dash-dot line is for an Einstein-de Sitter cosmological model ($\Omega_{m}=1.0$, $\Omega_{\Lambda} = 0.0$). In both models, we assume a Hubble constant of 70 km s$^{-1}$ Mpc$^{-1}$. We also present the number of SNe~II available at this epoch ($N_{\rm SNe}$), the epoch after the explosion, and the observed dispersion ($\sigma_{\rm obs}$).}
\label{fig:cosmo_sniip}
\end{figure*}

\begin{figure}
\centering
\includegraphics[width=1.0\columnwidth]{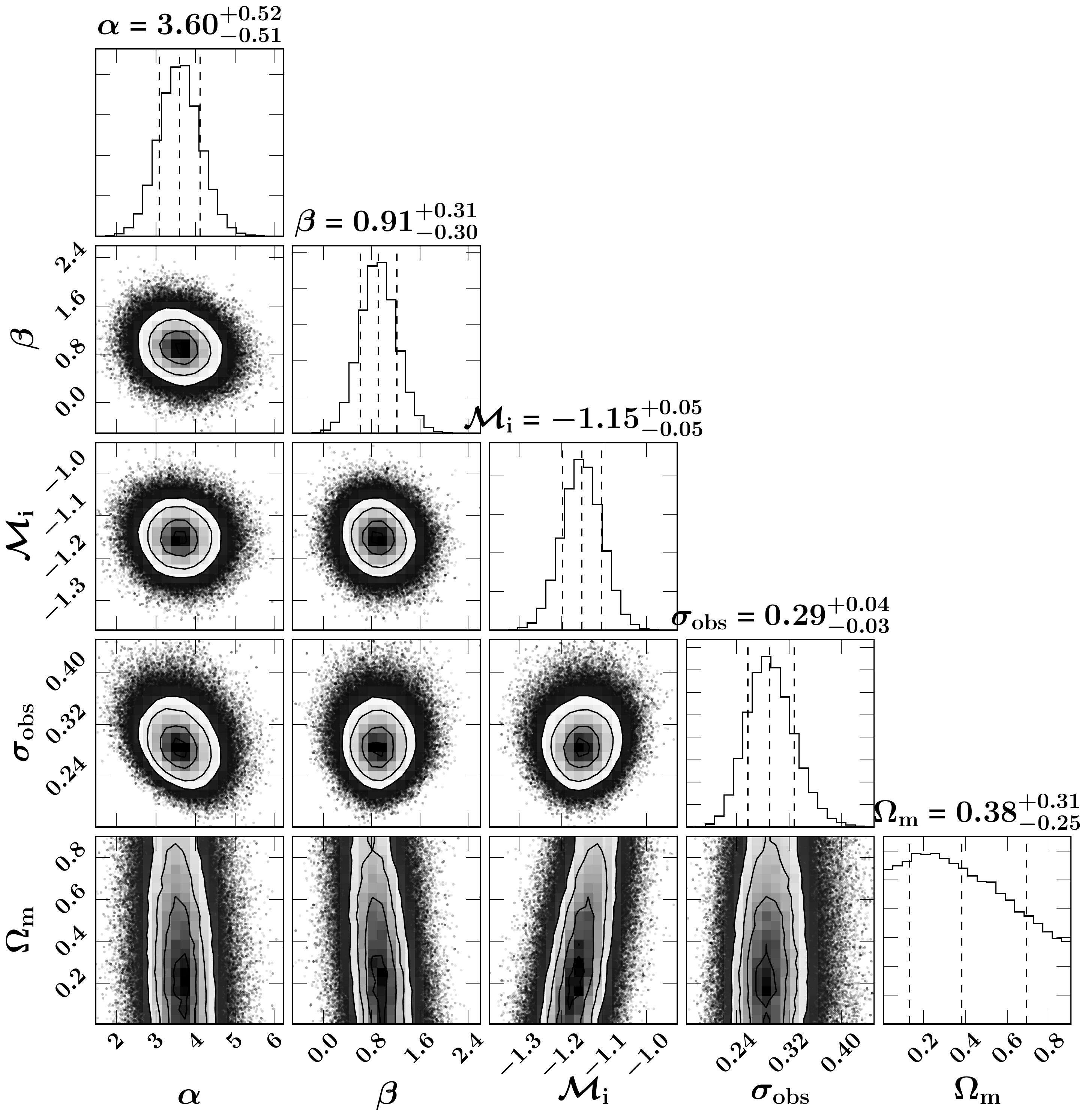}
\caption{Corner plot showing all of the one- and two-dimensional projections. Contours are shown at 1$\sigma$, 2$\sigma$, and 3$\sigma$. The five free parameters are plotted: $\alpha$, $\beta$, $\mathcal{M}_{\lambda 1}$, $\sigma_{\rm obs}$, and $\Omega_{m}$. To make this figure we use the corner plot package (triangle.py v0.1.1. Zenodo. 10.5281/zenodo.11020). In deriving this figure, we assume a flat universe.}
\label{fig:corner_plot}
\end{figure}

\subsection{Light-curve interpolation}\label{section_GP}

We model the SN light curves using the hierarchical Gaussian processes (also known as Bayesian smoothing splines) to interpolate the magnitudes and colours at different epochs. This technique has the advantage over other methods of allowing the inclusion of uncertainty information, thus producing less-biased interpolated values. Additionally, this method is very powerful for SN light curves having incomplete or noisy photometric data \citep{mandel09,mandel11,burns14,lochner16}. 

For this purpose, we use the fast and flexible Python library \textsc{George} developed by \citet{ambikasaran14}. In Figure \ref{fig:GP}, two examples of incomplete and noisy SN light-curve fits are shown. These two SNe (SN~2007nr and SN~05D4dn) are from the SDSS and SNLS samples, respectively. As shown in the figure, the best fit from the Gaussian process is very satisfactory during the plateau phase and allows us to derive the magnitudes and colours. Note that this procedure differs from that of \citet{dejaeger17a}, who did a simple linear interpolation. In this figure, we also clearly see the differences between the magnitudes derived using a linear or a Gaussian interpolation. For instance, for SN~2007nr at epoch 44 days after the explosion, the difference between the two methods in the $r$-band is $\sim 0.2$ mag. In addition to SN~2007nr and SN~05D4dn, five other SNe show differences in the $r$ or $i$ bands larger than 0.1 mag: SN~18321, SN~2007kz, SN~2007nv, SN~2007nw, and SN~2007ny (all from the SDSS). Finally, it is important to note that in this work, we never extrapolate the magnitude.

\subsection{Hubble diagram}

In the SCM, we use two corrections to standardise the SN observed magnitude: the expansion velocity, and the colour correction that accounts for host-galaxy extinction. Thus, the observed magnitude can be modeled as
\begin{ceqn}
\begin{align}
\begin{split}
m^\mathrm{model}_{i}=&\mathcal{M}_{i}-\alpha \mathrm{log_{10}}\left(\frac{v_\mathrm{H\beta}}{<v_\mathrm{H\beta}>~\mathrm{km~s^{-1}}}\right) \\ +& \beta (r-i) + 5 \mathrm{log_{10}} (\mathcal{D}_{L}(z_\mathrm{CMB}|\Omega_\mathrm{m},\Omega_\mathrm{\Lambda})),
\end{split}
\label{m_model}
\end{align}
\end{ceqn}
where $(r-i)$ is the colour, $\mathcal{D}_{L}(z_\mathrm{CMB}|\Omega_\mathrm{m},\Omega_\mathrm{\Lambda}$) is the luminosity distance ($\mathcal{D}_{L}$=H$_{0}$$d_{L}$) for a cosmological model depending on the cosmological parameters $\Omega_{m}$, $\Omega_{\Lambda}$, the CMB redshift $z_{\rm CMB}$, and the Hubble constant. Finally, $\alpha$, $\beta$, and $\mathcal{M}_{i}$ are also free parameters, with $\mathcal{M}_{i}$ corresponding to the ``Hubble-constant-free'' absolute magnitude ($\mathcal{M}_{i}$=M$_{i}$-5 log$_{10}$(H$_{0}$) + 25). Note that we center the velocity and colour ($(r-i)$-$<(r-i)>$) distributions using the mean velocity ($< v_{H\beta}>$ $\approx 5910$ km s$^{-1}$) and the mean colour ($<(r-i)> \approx -0.02$ mag) of the whole sample, respectively.

To derive the Hubble diagram and to determine the best-fit parameters, as done by \citet{poznanski09}, \citet{andrea10}, and \citet{dejaeger17a}, a Monte Carlo Markov Chain (MCMC) simulation is performed. In particular, here we use the Python package \textsc{EMCEE} developed by \citet{foreman13}, and we minimise the likelihood function defined as

\begin{ceqn}
\begin{align}
-2\mathrm{ln}(\mathcal{L})=\sum_\mathrm{SN} \left \{ \frac{\left [m^\mathrm{obs}_{i}- m^\mathrm{model}_{i} \right ]^{2}}{\sigma^{2}_\mathrm{tot}} +\mathrm{ln}(\sigma^{2}_\mathrm{tot}) \right \},
\label{likelihood}
\end{align}
\end{ceqn}
where we sum over all SNe~II available for one specific epoch, $m_{i}^\mathrm{obs}$ is the observed magnitude corrected for AKS, $m_{i}^\mathrm{model}$ is the model defined in Equation \ref{m_model}, and the total uncertainty $\sigma_\mathrm{tot}$ (corresponding to the error propagation of the model) is defined as
\begin{ceqn}
\begin{align}
\begin{split}
\sigma^{2}_\mathrm{tot}=&\sigma^{2}_{m_{i}} + (\frac{\alpha}{\mathrm{ln} 10}\frac{\sigma_{v_\mathrm{H\beta}}}{v_\mathrm{H\beta}})^2 + (\beta \sigma_{(r-i)})^2\\&+\left (\sigma_{z} \frac{5(1+z)}{z(1+z/2)\mathrm{ln}(10)}\right)^{2}+\sigma^{2}_\mathrm{obs} .
\end{split}
\end{align}
\end{ceqn}

\noindent Note that to measure the deviations between the observed SNe and the model, a term ($\sigma_\mathrm{obs}$) is added to the total error. This value includes true intrinsic scatter in the Hubble diagram (minimum uncertainty in any distance determination using the SCM) and any misestimates of the photometric, velocity, or redshift uncertainties. This term is also known as $\sigma_\mathrm{sys}$ \citep{poznanski09} or $\sigma_\mathrm{int}$ (e.g., \citealt{conley11,dejaeger17a}).

\section{Results}

In this results section, we will first attempt to extend the low-redshift Hubble diagram to higher redshifts, and then constrain the cosmological parameters.

\subsection{Fixed cosmology}\label{fixed_omega_m}

We use the complete SN~II sample, consisting of slowly declining (SNe~IIP) and rapidly declining (SNe~IIL) objects, as \citet{dejaeger15b} demonstrated that the Hubble residual does not depend on the slope of the plateau. This was also recently confirmed by \citet{gall17}. Our sample available at epoch 44 days after the explosion consists of 60 SNe~II (39\footnote{With respect to \citet{dejaeger17a}, SN~2005es and SN~2008F are not used owing to the lack of a spectrum at epochs later than 15 days (our cut for the cross-correlation method), while SN~2006ms is available at epoch 44 days after the explosion.}, 16, and 5 SNe~II from CSP, SDSS, and SNLS, respectively) together with the high-$z$ SN~2016jhj from HSC. The relevant information for all SNe~II in this sample is given in Appendix \ref{AppendixA}, Table \ref{SN_sample}. Note that the majority of the distance moduli are consistent with those of \citet{dejaeger17a}, but seven SNe~II (one CSP-I, four SDSS, and two SNLS) are not, because of differences in velocities (the cross-correlation method shows some bias for the extreme epochs; see \ref{txt:cc_method}) or in magnitudes/colours (some of these SNe~II have very noisy light curves, and were only linearly interpolated by \citet{dejaeger17a}).

First, we try to construct a high-$z$ Hubble diagram by finding the best-fit values ($\alpha$, $\beta$, $\mathcal{M}_{i}$, and $\sigma_{\rm obs}$) using a flat Universe ($\Omega_{m} + \Omega_{\Lambda} = 1$, $\Omega_{m}=0.3$) assumption. We find $\alpha = 3.57^{+0.52}_{-0.50}$, $\beta = 0.95^{+0.29}_{-0.29}$, and $\mathcal{M}_{i} = -1.16^{+0.04}_{-0.04}$, with an observed dispersion $\sigma_{\rm obs} = 0.27$ mag. In Figure \ref{fig:cosmo_sniip}, the Hubble diagram and the Hubble residuals of the combined data are displayed. Note that to derive the best epoch to use, the minimisation of dispersion in the Hubble diagram is our figure of merit. The best epoch is found to be 44 days after the explosion during the plateau phase. This epoch is similar to the 50 days in the rest frame (post-explosion) used by other SN~II cosmology studies \citep{nugent06,poznanski09,olivares10,andrea10}.

The observed dispersion found is consistent with the results from previous studies (0.26 mag, \citealt{nugent06}; 0.25 mag, \citealt{poznanski10}; 0.29 mag, \citealt{andrea10}; and 0.27 mag, \citealt{dejaeger17a}) and corresponds to 12--13\% in distance. Note that without any correction, the observed dispersion is more than twice as large (0.60 mag; $\sim 27$\% in distance). The best-fit parameters are nearly identical to those of \citet{dejaeger17a}, except for $\alpha$, which is slightly different ($\alpha = 3.57^{+0.52}_{-0.50}$ versus $\alpha = 3.18^{+0.41}_{-0.41}$). The discrepancy in $\alpha$ is easily explained by the difference of methodology used to derive the velocities. \citet{dejaeger17a} performed the minimum absorption technique, while in this work we use the cross-correlation method. If the cross-correlation method is applied only to the \citet{dejaeger17a} sample (without SN~2016jhj), we derive a value of $\alpha$ identical to that derived with the whole sample.

\subsection{$\Omega_{m}$ derivation}

As demonstrated in Section \ref{fixed_omega_m}, we are able to construct a high-$z$ Hubble diagram in which the differences between the expansion histories start to be distinguishable. Thus, as performed by \citet{dejaeger17a}, we try to put some constraints on the cosmological parameters. For this purpose, we assume a flat universe ($\Omega_{m} + \Omega_{\Lambda} = 1$), a Hubble constant $H_{0} = 70$ km s$^{-1}$ Mpc$^{-1}$, and leave $\Omega_{m}$ as a free parameter together with $\alpha$, $\beta$, $\mathcal{M}_{i}$, and $\sigma_{\rm obs}$. All of the best-fit parameters ($\alpha$, $\beta$, $\mathcal{M}_{i}$, and $\sigma_{\rm obs}$) are shown in Figure \ref{fig:corner_plot}, where a corner plot with all of the one- and two-dimensional projections is displayed. Note that all the parameters are slightly different from those obtained for a fixed cosmology ($\Lambda$CDM cosmological model) because here $\Omega_{m}$ is left as a free parameter.

A value for the matter density of $\Omega_{m} = 0.38^{+0.31}_{-0.25}$ is derived, which gives a density of dark energy of $\Omega_{\Lambda}=0.62^{+0.25}_{-0.31}$. Our prior probability distribution is defined to have uniform probability for $0.01 \leq\Omega_{m} \leq 0.9$, as well as $\alpha$, $\beta$, $\mathcal{M}_{\lambda 1}$ $\neq 0$. Note that if we choose a more restrictive prior for $\Omega_{m}$, such as $0.2 \leq \Omega_{m} \leq 0.6$, the uncertainties decrease and we derive a value of $\Omega_{m} = 0.38^{+0.14}_{-0.13}$. On the other hand, if we take less restrictive limits ($0.01 \leq \Omega_{m} \leq 2.5$,), the values and the uncertainties are not much larger ($\Omega_{m} = 0.50^{+0.56}_{-0.34}$). 

The value derived in this work is consistent with that obtained by \citet{dejaeger17a} ($\Omega_{m} = 0.41^{+0.31}_{-0.27}$) and confirms the evidence for dark energy using SNe~II. While the uncertainties we obtain here are far from the precision achieved using SNe~Ia (e.g., \citealt{betoule14}), this work confirms the great potential of SNe~II as distance indicators and our capacity to extend the current Hubble diagrams beyond $z=0.3$. Nevertheless, this work addresses the necessity of dedicating more observing time to high-redshift SNe~II in order to improve their utility as independent distance indicators and make them comparable in precision to SNe~Ia.

\section{Conclusions}

In this paper, we demonstrate that we are able to apply the SCM and extend the SN~II Hubble diagram beyond $z=0.3$. Although SNe~II are not currently competitive with SNe~Ia in terms of dispersion ($\sim 0.27$ mag vs. $\sim 0.10$ mag) or sample size ($\sim 61$ SNe~II vs. $\sim 740$ SNe~Ia), this work is comparable to the early SN~Ia results \citep{perlmutter97}, showing that SNe~II are a useful complementary and independent method for constraining the nature of dark energy. We summarise our results as follows.
\begin{enumerate}
\item{We test the cross-correlation method proposed by \citet{poznanski09} at high redshift and confirm its potential. This will be an asset for low-S/N spectra.}
\item{We obtain a dispersion of 0.27 mag using the SCM and 61 SNe~II at a redshift up to $\sim 0.34$, which is most distant SN~II Hubble diagram ever built using the SCM.}
\item{This high-$z$ Hubble diagram confirmed the result found in the literature in terms of dispersion or best-fit parameters ($\alpha$, $\beta$, $\mathcal{M}_{i}$).}
\item{We derived cosmological parameters ($\Omega_{m}$) consistent with $\Lambda$CDM: $\Omega_{m} = 0.38^{+0.31}_{-0.25}$. The uncertainties are somewhat better than those in previous studies of SN~II cosmology.}
\end{enumerate}

\section*{Acknowledgements}

We thank the referee for their thorough reading of the manuscript, which helped clarify and improve it. T.d.J. also thanks P. Nugent and M. Sullivan for helpful comments on this manuscript. Support for A.V.F.'s supernova research group at U.C. Berkeley have been provided by US NSF grant AST-1211916, the TABASGO Foundation, Gary and Cynthia Bengier (T.d.J. is a Bengier Postdoctoral Fellow), the Christopher R. Redlich Fund, and the Miller Institute for Basic Research in Science (U.C. Berkeley). The work of A.V.F. was completed in part at the Aspen Center for Physics, which is supported by NSF grant PHY-1607611; he thanks the Center for its hospitality during the neutron stars workshop in June and July 2017. L.G. was supported in part by the NSF under grant AST-1311862. S.G.G, M.H, and G.P acknowledge support from the Ministry of Economy, Development, and Tourism’s Millennium Science Initiative through grant IC120009, awarded to The Millennium Institute of Astrophysics (MAS). This work has been supported by Japan Society for the Promotion of Science (JSPS) KAKENHI Grant JP15H02075 (M.T.), JP25800103 (M.T.), JP26400222, JP16H02168, and JP17K05382 (K.N.). K.M acknowledges the JSPS Open Partnership Bilateral Joint Research Project between Japan and Chile, the YITP workshop YITP-T-16-05 supported by the Yukawa Institute for Theoretical Physics at Kyoto University, and JSPS KAKENHI Grant 17H02864. The work of the CSP-I has been supported by the NSF under grants AST-0306969, AST-0607438, and AST-1008343. 

This paper is based on data collected at the Subaru Telescope and retrieved from the HSC data archive system, which is operated by the Subaru Telescope and Astronomy Data Center at National Astronomical Observatory of Japan (NAOJ). The Hyper Suprime-Cam (HSC) collaboration includes the astronomical communities of Japan and Taiwan, and Princeton University. The HSC instrumentation and software were developed by the NAOJ, the Kavli Institute for the Physics and Mathematics of the Universe (Kavli IPMU), the University of Tokyo, the High Energy Accelerator Research Organization (KEK), the Academia Sinica Institute for Astronomy and Astrophysics in Taiwan (ASIAA), and Princeton University. Funding was contributed by the FIRST program from Japanese Cabinet Office, the Ministry of Education, Culture, Sports, Science and Technology (MEXT), the Japan Society for the Promotion of Science (JSPS), Japan Science and Technology Agency (JST), the Toray Science Foundation, NAOJ, Kavli IPMU, KEK, ASIAA, and Princeton University. 
The Pan-STARRS1 Surveys (PS1) have been made possible through contributions of the Institute for Astronomy, the University of Hawaii, the Pan-STARRS Project Office, the Max-Planck Society and its participating institutes, the Max Planck Institute for Astronomy, Heidelberg and the Max Planck Institute for Extraterrestrial Physics, Garching, The Johns Hopkins University, Durham University, the University of Edinburgh, Queen's University Belfast, the Harvard-Smithsonian Center for Astrophysics, the Las Cumbres Observatory Global Telescope Network Incorporated, the National Central University of Taiwan, the Space Telescope Science Institute, the National Aeronautics and Space Administration (NASA) under Grant No. NNX08AR22G issued through the Planetary Science Division of the NASA Science Mission Directorate, the NSF under grant AST-1238877, the University of Maryland, and Eotvos Lorand University (ELTE). This paper makes use of software developed for the Large Synoptic Survey Telescope. We thank the LSST Project for making their code available as free software at http://dm.lsst.org. 

Some of the data presented herein were obtained at the W.M. Keck Observatory, which is operated as a scientific partnership among the California Institute of Technology, the University of California, and NASA; the observatory was made possible by the generous financial support of the W.M. Keck Foundation. This work is based in part on data produced at the Canadian Astronomy Data Centre as part of the CFHT Legacy Survey, a collaborative project of the National Research Council of Canada and the French Centre National de la Recherche Scientifique. The work is also based on observations obtained at the Gemini Observatory, which is operated by the Association of Universities for Research in Astronomy, Inc., under a cooperative agreement with the NSF on behalf of the Gemini partnership: the NSF, the STFC (United Kingdom), the National Research Council (Canada), CONICYT (Chile), the Australian Research Council (Australia), CNPq (Brazil) and CONICET (Argentina). This research used observations from Gemini program number: GN-2005A-Q-11, GN-2005B-Q-7, GN-2006A-Q-7, GS-2005A-Q-11 and GS-2005B-Q-6, and GS-2008B-Q-56. This research has made use of the NASA/IPAC Extragalactic Database (NED), which is operated by the Jet Propulsion Laboratory, California Institute of Technology, under contract with NASA and of data provided by the Central Bureau for Astronomical Telegrams.




\appendix

\section{}\label{AppendixA}

In Table \ref{SN_sample}, the relevant information for all SNe~II used in the Hubble diagram is displayed. The first column gives the SN name, followed (in Column 2) by its reddening due to dust in our Galaxy \citep{schlafly11}. In Column 3, we list the host-galaxy velocity in the CMB frame using the CMB dipole model presented by \citet{fixsen96}. The explosion epoch is given in Column 4. In Column 5, the magnitude in the $i$ band at epoch 44 days post-explosion is listed, followed by the $r-i$ colour at the same epoch in Column 6. Column 7 gives the H$\beta$ velocity at epoch 44 days. Finally, in Columns 8 and 9 we respectively present the distance modulus measured using SCM and the survey from which the SN~II originates.
\begin{table*}
\centering
\caption{The supernova sample}
\begin{tabular}{lccccccccc}
\hline
\hline
SN & $A_V$G & $z_{\rm CMB}$ (err) &Explosion date & $m_{i}$ & $r-i$ & $v_{{\rm H}\beta}$  &$\mu_{\rm SCM}$ & Campaign\\  & mag &  & MJD &mag &mag & km s$^{-1}$ & mag &\\
\hline
SN04er & 0.070 & 0.014 (0.0001) & 53271.8 (4.0) & 16.72 (0.01) & 0.208 (0.016) & 7504 (453) & 33.88 (0.13) & CSP-I\\
SN05J & 0.075 & 0.015 (0.0001) & 53382.8 (7.0) & 16.99 (0.01) & -0.031 (0.011) & 6039 (273) & 33.99 (0.10) & CSP-I\\
SN05K & 0.108 & 0.028 (0.0001) & 53369.8 (7.0) & 18.81 (0.02) & -0.127 (0.023) & 5517 (831) & 35.75 (0.30) & CSP-I\\
SN05Z & 0.076 & 0.020 (0.0001) & 53396.7 (8.0) & 17.46 (0.01) & 0.087 (0.014) & 7037 (288) & 34.61 (0.09) & CSP-I\\
SN05an & 0.262 & 0.012 (0.0001) & 53426.7 (4.0) & 16.74 (0.01) & -0.021 (0.012) & 6321 (368) & 33.81 (0.12) & CSP-I\\
SN05dk & 0.134 & 0.015 (0.0001) & 53599.5 (6.0) & 16.77 (0.02) & -0.069 (0.027) & 6488 (474) & 33.92 (0.15) & CSP-I\\
SN05dt & 0.079 & 0.025 (0.0001) & 53605.6 (9.0) & 18.55 (0.01) & 0.043 (0.018) & 4824 (404) & 35.12 (0.17) & CSP-I\\
SN05dw & 0.062 & 0.017 (0.0001) & 53603.6 (9.0) & 17.67 (0.01) & -0.001 (0.02) & 5469 (537) & 34.49 (0.20) & CSP-I\\
SN05dx & 0.066 & 0.026 (0.0001) & 53615.9 (7.0) & 19.24 (0.05) & 0.283 (0.059) & 4382 (328) & 35.45 (0.17) & CSP-I\\
SN05dz & 0.223 & 0.018 (0.0001) & 53619.5 (4.0) & 17.92 (0.02) & 0.003 (0.028) & 5665 (442) & 34.79 (0.16) & CSP-I\\
SN05lw & 0.135 & 0.027 (0.0001) & 53716.8 (5.0) & 17.96 (0.02) & 0.167 (0.024) & 7037 (480) & 35.05 (0.14) & CSP-I\\
SN05me & 0.070 & 0.022 (0.0001) & 53721.6 (6.0) & 18.34 (0.01) & -0.059 (0.011) & 5869 (482) & 35.32 (0.17) & CSP-I\\
SN06Y & 0.354 & 0.034 (0.0001) & 53766.5 (4.0) & 18.66 (0.03) & -0.137 (0.036) & 6853 (348) & 35.96 (0.11) & CSP-I\\
SN06ai & 0.347 & 0.015 (0.0001) & 53781.8 (5.0) & 16.82 (0.02) & -0.077 (0.026) & 6226 (333) & 33.9 (0.12) & CSP-I\\
SN06bl & 0.144 & 0.033 (0.0001) & 53823.8 (6.0) & 18.12 (0.01) & -0.064 (0.016) & 6168 (341) & 35.18 (0.12) & CSP-I\\
SN06ee & 0.167 & 0.014 (0.0001) & 53961.9 (4.0) & 17.47 (0.02) & 0.005 (0.025) & 3418 (297) & 33.50 (0.18) & CSP-I\\
SN06ms & 0.095 & 0.015 (0.0001) & 54034.0 (12.0) & 17.75 (0.02) & 0.004 (0.029) & 4191 (835) & 34.12 (0.39) & CSP-I\\
SN06qr & 0.126 & 0.015 (0.0001) & 54062.8 (7.0) & 18.12 (0.01) & 0.075 (0.014) & 4535 (514) & 34.56 (0.23) & CSP-I\\
SN07P & 0.111 & 0.042 (0.0001) & 54118.7 (3.0) & 18.96 (0.02) & -0.110 (0.023) & 6143 (585) & 36.06 (0.19) & CSP-I\\
SN07U & 0.145 & 0.026 (0.0001) & 54134.6 (6.0) & 17.73 (0.01) & -0.091 (0.019) & 6863 (381) & 34.99 (0.12) & CSP-I\\
SN07W & 0.141 & 0.011 (0.0001) & 54136.8 (7.0) & 17.31 (0.01) & -0.014 (0.017) & 3441 (345) & 33.37 (0.20) & CSP-I\\
SN07ab & 0.730 & 0.024 (0.0001) & 54123.8 (6.0) & 17.63 (0.02) & -0.062 (0.024) & 7838 (449) & 35.09 (0.12) & CSP-I\\
SN07hm & 0.172 & 0.024 (0.0001) & 54335.6 (6.0) & 18.77 (0.01) & -0.074 (0.016) & 6161 (313) & 35.84 (0.11) & CSP-I\\
SN07il & 0.129 & 0.021 (0.0001) & 54349.8 (4.0) & 17.8 (0.02) & -0.027 (0.022) & 6168 (337) & 34.83 (0.12) & CSP-I\\
SN07sq & 0.567 & 0.016 (0.0001) & 54421.8 (3.0) & 17.84 (0.01) & 0.369 (0.016) & 7167 (476) & 34.79 (0.14) & CSP-I\\
SN08W & 0.267  & 0.020 (0.0001) & 54485.8 (6.0) & 17.91 (0.02) & 0.053 (0.029) & 5620 (347) & 34.72 (0.13) & CSP-I\\
SN08ag & 0.229 & 0.015 (0.0001) & 54479.8 (6.0) & 16.82 (0.01) & -0.033 (0.018) & 4795 (318) & 33.44 (0.14) & CSP-I\\
SN08aw & 0.111 & 0.011 (0.0001) & 54517.8 (10.0) & 16.11 (0.01) & 0.015 (0.016) & 6704 (261) & 33.24 (0.09) & CSP-I\\
SN08bh & 0.060 & 0.015 (0.0001) & 54543.5 (5.0) & 17.92 (0.01) & 0.194 (0.015) & 6201 (478) & 34.77 (0.16) & CSP-I\\
SN08br & 0.255 & 0.011 (0.0001) & 54555.7 (9.0) & 17.81 (0.01) & 0.103 (0.016) & 2608 (507) & 33.32 (0.38) & CSP-I\\
SN08bu & 1.149 & 0.022 (0.0001) & 54566.8 (5.0) & 18.32 (0.03) & 0.288 (0.045) & 5517 (586) & 34.90 (0.22) & CSP-I\\
SN08ga & 1.865 & 0.015 (0.0001) & 54711.8 (4.0) & 17.18 (0.02) & -0.049 (0.022) & 5714 (368) & 34.11 (0.13) & CSP-I\\
SN08gi & 0.181 & 0.024 (0.0001) & 54742.7 (9.0) & 17.87 (0.01) & -0.006 (0.016) & 5949 (518) & 34.83 (0.18) & CSP-I\\
SN08gr & 0.039 & 0.022 (0.0001) & 54766.5 (4.0) & 17.41 (0.01) & -0.084 (0.016) & 7259 (389) & 34.76 (0.12) & CSP-I\\
SN08hg & 0.050 & 0.018 (0.0001) & 54779.7 (5.0) & 18.48 (0.02) & 0.069 (0.025) & 4406 (642) & 34.87 (0.29) & CSP-I\\
SN08if & 0.090 & 0.013 (0.0001) & 54807.8 (5.0) & 16.47 (0.02) & -0.136 (0.028) & 6805 (283) & 33.75 (0.10) & CSP-I\\
SN09ao & 0.106 & 0.012 (0.0001) & 54890.7 (4.0) & 16.93 (0.01) & 0.340 (0.013) & 5436 (346) & 33.44 (0.13) & CSP-I\\
SN09bu & 0.070 & 0.011 (0.0001) & 54907.9 (6.0) & 16.95 (0.01) & 0.132 (0.009) & 5619 (381) & 33.70 (0.14) & CSP-I\\
SN09bz & 0.110 & 0.011 (0.0001) & 54915.8 (4.0) & 16.91 (0.01) & -0.074 (0.012) & 5647 (386) & 33.83 (0.14) & CSP-I\\
8321   & 0.080 &0.107 (0.001) & 54353.6 (5.0) & 21.25 (0.04) & -0.479 (0.069) & 6838 (354) & 38.83 (0.13) & SDSS\\
SN06gq & 0.096 & 0.069 (0.0005) & 53992.4 (3.0) & 20.39 (0.02) & -0.143 (0.032) & 4708 (725) & 37.08 (0.31) & SDSS\\
SN06iw & 0.137 & 0.030 (0.0005) & 54010.7 (1.0) & 18.73 (0.01) & 0.061 (0.017) & 6300 (317) & 35.72 (0.11) & SDSS\\
SN06jl & 0.504 & 0.055 (0.0005) & 54006.8 (15.0) & 19.46 (0.02) & 0.053 (0.033) & 6450 (410) & 36.50 (0.14) & SDSS\\
SN06kn & 0.194 & 0.119 (0.0005) & 54007.0 (1.5) & 21.19 (0.08) & -0.067 (0.106) & 6218 (431) & 38.27 (0.19) & SDSS\\
SN06kv & 0.080 & 0.062 (0.0005) & 54016.5 (4.0) & 20.26 (0.05) & -0.072 (0.072) & 5208 (383) & 37.05 (0.17) & SDSS\\
SN07kw & 0.074 & 0.067 (0.0005) & 54361.6 (2.5) & 19.96 (0.02) & -0.055 (0.027) & 5854 (429) & 36.93 (0.15) & SDSS\\
SN07ky & 0.105 & 0.073 (0.0005) & 54363.5 (3.0) & 20.70 (0.03) & -0.080 (0.043) & 5069 (356) & 37.45 (0.15) & SDSS\\
SN07kz & 0.320 & 0.127 (0.0005) & 54362.6 (3.5) & 21.54 (0.06) & -0.239 (0.092) & 5981 (354) & 38.70 (0.16) & SDSS\\
SN07lb & 0.496 & 0.038 (0.0005) & 54368.8 (7.0) & 18.58 (0.01) & 0.022 (0.013) & 7550 (408) & 35.91 (0.12) & SDSS\\
SN07ld & 0.255 & 0.027 (0.005) & 54369.6 (5.5) & 18.19 (0.01) & -0.044 (0.009) & 6264 (367) & 35.26 (0.43) & SDSS\\
SN07lj & 0.118 & 0.049 (0.005) & 54370.2 (3.5) & 19.69 (0.02) & -0.055 (0.027) & 5796 (397) & 36.64 (0.27) & SDSS\\
SN07lx & 0.120 & 0.056 (0.0005) & 54374.5 (8.0) & 20.18 (0.03) & -0.039 (0.044) & 5282 (413) & 36.97 (0.17) & SDSS\\
SN07nr & 0.079 & 0.139 (0.0005) & 54353.5 (5.0) & 22.20(0.12) & -0.383 (0.152) & 5216 (327) & 39.25 (0.22) & SDSS\\
SN07nw & 0.204 & 0.056 (0.0005) & 54372.2 (7.0) & 20.42 (0.03) & 0.042 (0.048) & 6419 (736) & 37.46 (0.23) & SDSS\\
SN07ny & 0.080 & 0.142 (0.0005) & 54367.7 (7.0) & 21.81 (0.11) & -0.276 (0.162) & 6379 (377) & 39.11 (0.21) & SDSS\\
04D1pj & 0.076 & 0.155 (0.0005) & 53304.0 (8.0) & 22.39 (0.04) & -0.052 (0.049) & 6960 (332) & 39.64 (0.12) & SNLS\\
04D4fu & 0.072 & 0.132 (0.0005) & 53213.0 (6.0) & 22.36 (0.04) & -0.100 (0.04) & 6166 (329) & 39.46 (0.12) & SNLS\\
05D4dn & 0.073 & 0.190 (0.0005) & 53605.0 (7.0) & 23.41 (0.08) & -0.079 (0.089) & 5717 (860) & 40.36 (0.32) & SNLS\\
06D1jx & 0.079 & 0.134 (0.001) & 54068.0 (6.0) & 22.22 (0.02) & -0.088 (0.024) & 5876 (392) & 39.23 (0.14) & SNLS\\
06D2bt & 0.051 & 0.079 (0.001) & 53745.0 (10.0) & 20.95 (0.03) & -0.078 (0.05) & 5877 (416) & 37.95 (0.16) & SNLS\\
SN2016fvh & 0.052 & 0.341 (0.001) &57719.6 (2.0) & 23.32 (0.04) & -0.162 (0.041) & 9030 (452) & 41.07 (0.12) & HSC\\

\hline
\hline
\end{tabular}
\label{SN_sample}
\end{table*}

\bsp	
\label{lastpage}
\end{document}